\documentclass[iop,apj,letterpaper]{emulateapj}

\usepackage{apjfonts}
\usepackage{amsmath, amstext}
\usepackage{url}
\usepackage[backref, breaklinks, colorlinks=true, urlcolor=blue, anchorcolor=blue, linkcolor=blue, citecolor=blue,]{hyperref}
\usepackage[all]{hypcap}
\usepackage{bm}
\usepackage[normalem]{ulem}

\slugcomment{Draft version \today}

\shorttitle{Results from EDGES High-Band: I}
\shortauthors{Monsalve et al.}

\begin{document}

\title{Results from EDGES High-Band:\\I. Constraints on Phenomenological Models for the Global $21$ cm Signal}

\author{
Raul A. Monsalve\altaffilmark{1,2},
Alan E. E. Rogers\altaffilmark{3},
Judd D. Bowman\altaffilmark{2}, 
and Thomas J. Mozdzen\altaffilmark{2}
}

\affil{$^1$Center for Astrophysics and Space Astronomy, University of Colorado, Boulder, CO 80309, USA; \href{mailto:Raul.Monsalve@colorado.edu}{Raul.Monsalve@colorado.edu}}
\affil{$^2$School of Earth and Space Exploration, Arizona State University, Tempe, AZ 85287, USA}
\affil{$^3$Haystack Observatory, Massachusetts Institute of Technology, Westford, MA 01886, USA}

\begin{abstract}

We report constraints on the global $21$~cm signal due to neutral hydrogen at redshifts $14.8 \geq z \geq 6.5$.  We derive our constraints from low foreground observations of the average sky brightness spectrum conducted with the EDGES High-Band instrument between September~$7$ and October~$26$, $2015$. Observations were calibrated by accounting for the effects of antenna beam chromaticity, antenna and ground losses, signal reflections, and receiver parameters.  We evaluate the consistency between the spectrum and phenomenological models for the global $21$ cm signal. For tanh-based representations of the ionization history during the epoch of reionization, we rule out, at $\geq2\sigma$ significance, models with duration of up to $\Delta z = 1$ at $z\approx8.5$ and higher than $\Delta z = 0.4$ across most of the observed redshift range under the usual assumption that the $21$~cm spin temperature is much larger than the temperature of the cosmic microwave background (CMB) during reionization.  We also investigate a `cold' IGM scenario that assumes perfect Ly$\alpha$ coupling of the $21$~cm spin temperature to the temperature of the intergalactic medium (IGM), but that the IGM is not heated by early stars or stellar remants.  Under this assumption, we reject tanh-based reionization models of duration $\Delta z \lesssim 2$ over most of the observed redshift range.  Finally, we explore and reject a broad range of Gaussian models for the $21$~cm absorption feature expected in the First Light era. As an example, we reject $100$ mK Gaussians with duration (full width at half maximum) $\Delta z \leq 4$ over the range $14.2\geq z\geq 6.5$ at $\geq2\sigma$ significance. 

\end{abstract}

\keywords{early universe --- cosmology: observations --- methods: data analysis}

\section{Introduction}

Illuminating the early stellar history is important for understanding the origin and evolution of structure in the Universe.  The first luminous sources emitted UV and X-ray radiation that ionized and heated the diffuse neutral hydrogen gas in the intergalactic medium (IGM). Stars are believed to be responsible for the bulk of the UV photons, while the primary sources of X-rays are still largely uncertain, but usually assumed to be stellar remnants \citep{barkana2001, furlanetto2006b, pritchard2012, mesinger2013, madau2015, fialkov2016b, greig2017}.

Low frequency radio observations of the redshifted 21 cm line of neutral hydrogen gas represent a unique avenue for observationally constraining the radiative properties of the first luminous objects and the reionization history of the Universe \citep{madau1997, furlanetto2006b}. In particular, the sky average or `global' $21$ cm signal is expected to provide essential information through the mean properties of the IGM at $z\gtrsim6$ \citep{shaver1999, gnedin2004, mirocha2013, mirocha2015, fialkov2016a}. The brightness temperature of the global $21$ cm signal is modeled as \citep{zaldarriaga2004}
\begin{equation}
T_{21}(z) \approx 28~x_{\text{HI}}(z)\left[\frac{T_s(z) - T_{\text{cmb}}(z)} {T_s(z)}\right]\sqrt{\frac{1+z}{10}}\;\;\;\text{[mK]},
\label{equation_global_signal}
\end{equation}

\noindent where $T_{\text{cmb}}$ is the temperature of the cosmic microwave background (CMB), $x_{\text{HI}}$ is the average fraction of neutral hydrogen in the IGM, and $T_s$ is the spin temperature of the neutral hydrogen, which quantifies the relative abundance of hydrogen atoms in the high and low energy ground state.  Information about the radiative properties of the first stars, galaxies, and black holes is encapsulated in the neutral fraction and spin temperature of the gas and, therefore, can be probed through measurements of the global $21$~cm brightness temperature. Since redshift maps to frequency for the 21~cm line according to $\nu=1420\;\text{MHz}/(1+z)$, the brightness temperature for redshifts $\gtrsim6$ is measured at frequencies $\lesssim 200$~MHz in the VHF radio band.
 
Most models for the global 21~cm signal contain two dominant features once star formation commences \citep{shaver1999, gnedin2004, furlanetto2006a, mesinger2013, sitwell2014, tanaka2016, kaurov2016, fialkov2016b, mirocha2017, cohen2017}. First, the models predict an approximately Gaussian-shaped absorption trough during the First Light era. This trough is initiated at $z\gtrsim15$ with the onset of coupling of the spin temperature to the cold kinetic temperature of the IGM, $T_{\text{IGM}}$, by UV photons from the first sources \citep{wouthuysen1952, field1958}. The trough usually concludes with the subsequent heating of the IGM above the CMB temperature by X-rays from the early stellar remnants in the models. The second global 21~cm spectral feature is expected during the epoch of reionization (EoR, $z \lesssim 10$), when neutral hydrogen---and therefore the $21$~cm signal---is gradually extinguised by UV radiation from the increasing number of sources.  The evolution of the neutral fraction in the models during reionization often exhibits a functional form similar to a hyperbolic tangent.  

This general scenario for the evolution of the IGM and the global 21~cm signal is consistent among many theoretical models.  However, because the specific properties of early stars and stellar evolution remain unknown, the timing, duration, and amplitude of the 21~cm features all vary between models.  In most cases, for models based on stellar and galactic properties extrapolated from existing observations, reionization occurs after the IGM has been heated well above the CMB temperature and the 21~cm signal is in emission.  In some cases, however, reionization takes place without substantial heating and while the 21~cm signal is still in absorption.  Typically, the absorption trough and reionization transition in the global 21~cm signal have expected amplitudes of tens to hundreds of mK.     

The Experiment to Detect the Global EoR Signature (EDGES) aims to detect the global $21$ cm signal through single antenna radio observations. It has placed the most significant lower limit on the duration of reionization to date, of $\Delta z>0.06$ at $95$\% confidence \citep{bowman2010}.   Ongoing developments in instrumentation and calibration have improved the performance of the system, as well as enabled EDGES to contribute to the characterization of astronomical foregrounds and Earth's ionosphere \citep{rogers2008, rogers2012, rogers2015, mozdzen2016, monsalve2016, mozdzen2017, monsalve2017}.   EDGES currently operates two total power spectral radiometers, identified as Low-Band and High-Band, which cover the ranges $50-100$ MHz ($27.4\gtrsim z\gtrsim13.2$) and $90-190$ MHz ($14.8\gtrsim z\gtrsim6.5$), and nominally target the features from First Light and the reionization, respectively. They observe from the Murchison Radio-astronomy Observatory (MRO) in Western Australia, taking advantage of its radio-quiet environment \citep{bowman2010b, offringa2015}. 

Several other experiments are also pursuing measurement of the global 21~cm signal.  Using $4.4$ hours of data, SCI-HI reported limits in the form of $1$ K rms residuals in the range $60-88$ MHz after subtracting the foregrounds using a three term log-log polynomial \citep{voytek2014}.  LEDA, using $19$ minutes of effective measurements, placed limits on the absorption trough between $50$ and $100$ MHz in the form of $95$\% constraints on the amplitude ($>-890$ mK) and $1\sigma$ width ($>6.5$ MHz) of a Gaussian model for the trough \citep{bernardi2016}. The SARAS 2 experiment \citep{singh2017} has recently ruled out at $1\sigma$ several models of the global 21~cm signal during the reionization era that were generated from the semi-numerical simulations of \citet{cohen2017}. Finally, the DARE space mission is planning to measure the global $21$ cm signal from orbit above the far side of the Moon \citep{burns2017}.

In this paper we present constraints on the global $21$ cm signal from measurements conducted with EDGES High-Band between September~$7$ and October~$26$, $2015$. The data consist of $40$ nighttime observations over the low foreground region $0.26-6.26$ hr local sidereal time (LST). They are processed through a pipeline that performs instrumental calibration and excision of data contaminated with radio-frequency interference (RFI). The data are then averaged in time and binned in frequency to obtain the final sky temperature spectrum.

We use the average spectrum to probe phenomenological models for the global $21$ cm signal. Specifically, we probe the allowed duration of reionization under two end-member cases of the IGM thermal evolution.  Our two thermal cases are: (1) the standard `hot'  IGM scenario where early stars and stellar remants have heated the IGM before reionization such that the gas kinetic temperature is $T_{\text{IGM}}\gg T_{\text{cmb}}$, and (2) a `cold' IGM with no heating from stars or stellar remants before or during reionization, such that the gas kinetic temperature history is given by the adiabatic expansion of the Universe.  In both cases we assume $T_s=T_{\text{IGM}}$ \citep{ciardi2003, pritchard2007} and we represent the ionization history ($x_{\text{HI}}$) with functional forms drawn from hyperbolic tangents. We also conduct a general test of absorption signals expected from First Light using models of absorption trough features based on a broad set of symmetrical and skewed Gaussians. For all of our phenomenological cases, our full measurement model includes polynomial terms to account for foregrounds and calibration systematics, in addition to the $21$ cm contribution. We estimate our sensitivity to calibration uncertainties by propagating uncertainty estimates through the analysis pipeline. 

This paper is organized as follows. Section~\ref{section_instrument} describes the EDGES High-Band instrument and calibration. Section~\ref{section_data_analysis} presents the data analyzed and the strategy for constraining phenomenological parameters. Section~\ref{section_results} presents the main results of this paper, in the form of constraints on the tanh and Gaussian models, and discusses the impact of conservative calibration uncertainties. Section~\ref{section_conclusion} concludes this work with a summary of the findings.

\section{Instrument}
\label{section_instrument}
Here we describe the EDGES High-Band instrument, with a focus on the antenna characteristics and calibration.  The EDGES High-Band instrument captures sky radiation using a single polarization dipole-like blade antenna \citep{mozdzen2016, mozdzen2017}. This antenna is composed of two rectangular aluminum panels of dimensions $62.5$~cm $\times$ $48.1$~cm, mounted horizontally at a height of $52$ cm above a metal ground plane. The separation between the panels is $2.2$ cm. The panels are supported by a lightweight structure composed of hollow square fiberglass pipes and teflon rods. The metal ground plane rests on the physical ground and consists of a solid aluminum central square of $5.35$ m $\times$ $5.35$ m and four wiregrid panels of size $5$ m $\times$ $2$ m attached to the sides of the solid square. The excitation axis of the antenna is orientated $-5^{\circ}$ from the North-South axis.  The antenna beam points at the zenith and has a full width at half maximum (FWHM) at $140$ MHz of $72^{\circ}$ ($108^{\circ}$) parallel (perpendicular) to the excitation axis. 

The front-end receiver is located under the ground plane, directly below the antenna. A Roberts balun \citep{roberts1957} is used to ground reference the electrically balanced signal induced at the antenna panels and guide it toward the receiver. The balun is implemented as two brass tubes of 0.5" outer diameter connected to the antenna panels, and a copper plated brass rod running up inside one of the tubes. A copper plate connects the rod sticking out of the tube above the antenna panel, with the opposite panel. At the base of the balun the tubes are attached to the ground plane. A small rectangular metal enclosure is also attached to the ground plane to shield against vertical currents in the tubes. The tube containing the rod extends below the ground plane, where it transitions into a teflon dielectric SMA connector that takes the signal to the receiver input.

The front-end receiver is designed around a low-noise amplifier (LNA), plus additional stages of gain, filtering, and conditioning. The input of the LNA switches continuously between: (1) the antenna, (2) an ambient load noise reference, and (3) the ambient load plus an active noise source connected in series to form a second, higher noise reference. The measurements of these two references are used to conduct a relative calibration at the LNA input, which removes the time dependent instrument passband. The duration of the three-position switching cycle is $39$ seconds. A back-end amplification unit, located $\approx100$ m away from the antenna/receiver, provides additional stages of gain, filtering, and conditioning. The output from the back-end is digitized and Fourier transformed at a resolution of $6.1$ kHz. We store the power spectral densities from the antenna and the two internal noise references for offline calibration.

The reflection coefficient of the antenna, including the effect of the balun, is a critical parameter in the EDGES calibration approach \citep{rogers2012, monsalve2017}. This parameter is measured in the field using a vector network analyzer (VNA), located next to the back-end unit, in coordination with electronics at the front-end receiver. The measurement is conducted remotely, without disconnecting the antenna from the receiver.

\subsection{Calibration}
\label{section_calibration}

To convert the antenna power spectral density measured in the field to a calibrated sky temperature spectrum, we use  calibration measurements and models to remove the effects of the following parameters: (1) gain, offset, reflection coefficient, and noise parameters of the receiver, (2) antenna reflection coefficient, (3) antenna and ground losses, and (4) antenna beam chromaticity.

\subsubsection{Receiver}

We set the input of the receiver as the reference plane for absolute antenna temperature measurements. We calibrate the sky measurements at this plane using six receiver parameters obtained in the laboratory before deployment \citep{rogers2012, monsalve2016, monsalve2017}.   During laboratory calibration and operation in the field, the receiver is maintained at a nominal temperature of $25^{\circ}$C using active control.  During the nighttime sky measurements used for this paper, the receiver temperature drifted within $\pm1^{\circ}$C. To remove the effect of these drifts, we conducted an additional laboratory calibration with the receiver temperature set to $35^{\circ}$C.  We use the two laboratory results to calibrate the sky measurements using receiver parameters interpolated (or extrapolated) linearly to the measured temperature of the receiver.

\subsubsection{Antenna Reflection Coefficient}
\label{section_antenna_reflection_coefficient}

We measured in the field the reflection coefficient of the antenna, including the effect of the balun, during a three-day session starting on September 19, 2015, at a rate of one measurement per minute. The observed nighttime and night-to-night reflection stability is better than $\pm0.01$ dB and $\pm0.1^{\circ}$. This stability is comparable to the intrinsic uncertainty in the VNA reflection measurement \citep{monsalve2016, monsalve2017}. Therefore, we average all $29$~hours of nighttime measurements from the session to derive our fiducial antenna reflection coefficient.

\begin{figure}
\centering
\includegraphics[width=0.49\textwidth]{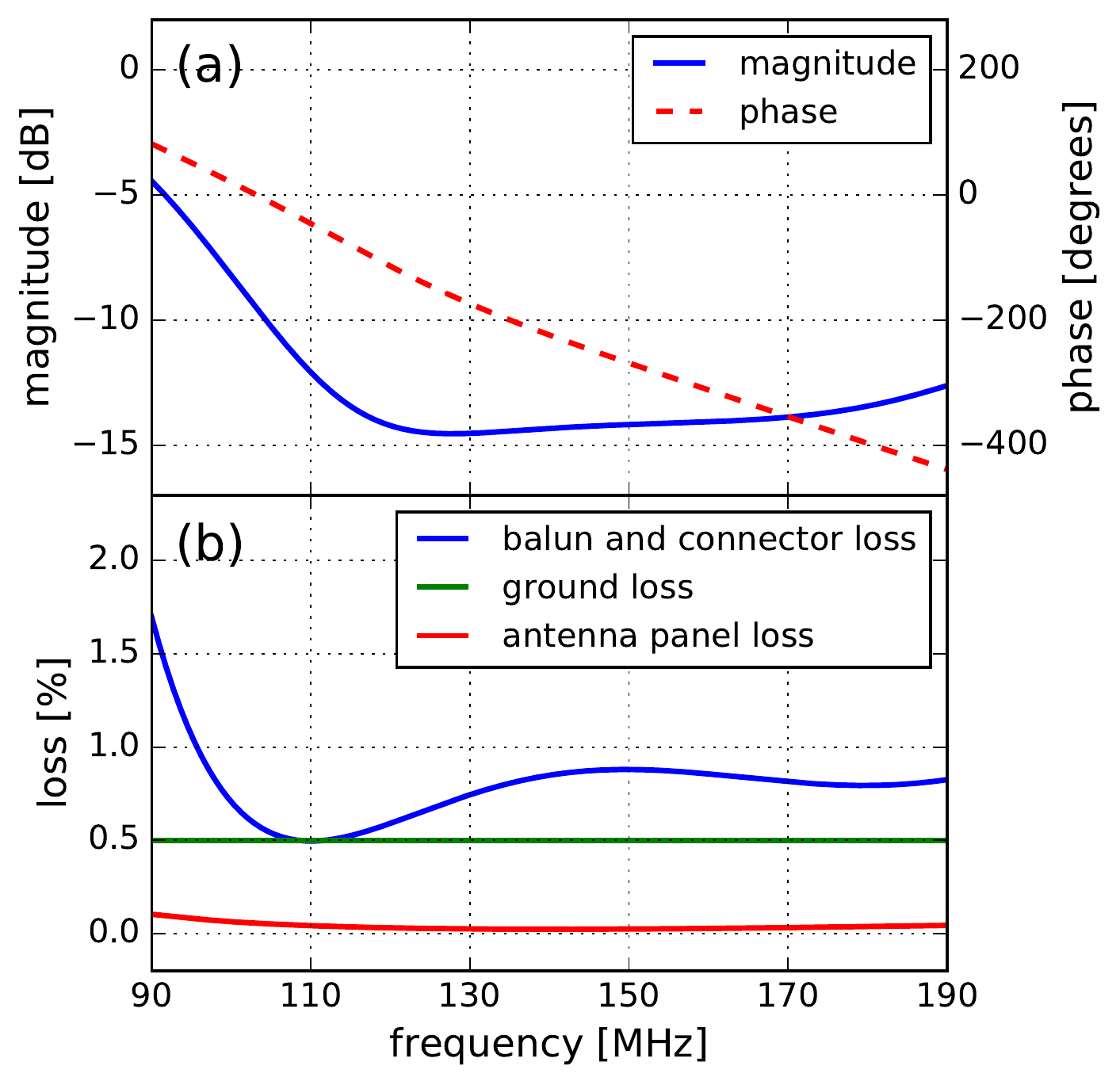}
\caption{(a) Polynomial fits to the measured antenna reflection coefficient magnitude and phase as discussed in Section~\ref{section_antenna_reflection_coefficient}. (b) Fiducial antenna and ground losses described in Section~\ref{section_antenna_losses} and used to correct the measured spectra.}
\label{figure_antenna_losses_s11}
\end{figure}

We fit the reflection magnitude and phase using $16$ term polynomials in frequency in order to remove noise and interpolate the reflection to the same frequency channels as the sky spectra. These models capture all significant structure in the data and produce rms fit residuals of $0.004$ dB and $0.03^{\circ}$.  Figure \ref{figure_antenna_losses_s11} (a) shows the magnitude and phase of the antenna reflection coefficient.

\subsubsection{Antenna and Ground Losses}
\label{section_antenna_losses}

After calibrating the antenna temperature at the receiver input, we need to remove from the spectrum the effect of losses in the antenna panels ($L_a$), the balun ($L_b$) and connector to the receiver ($L_c$), as well as ground losses due to non zero beam directivity below the horizon ($L_g$). The measured antenna temperature affected by losses, $T^L_{\text{ant}}$, is given by

\begin{equation}
T^L_{\text{ant}} = L T_{\text{ant}} + \left(1-L\right) T_{\text{amb}},
\label{equation_loss}
\end{equation}

\noindent where $T_{\text{ant}}$ is the antenna temperature before losses, $T_{\text{amb}}$ is the ambient temperature, and $L=L_aL_bL_cL_g$ represents the combined loss, which takes values between $0$ and $1$ where $1$ represents no loss. 

The antenna panels are affected by resistive losses. We estimate this small effect from electromagnetic simulations with the FEKO software package\footnote{\url{www.feko.info}}. Balun and connector loss occurs along the conductors, as well as through their air and teflon dielectric, respectively. We estimate these losses from analytical models of these cylindrical transmission lines, and check these models using reflection measurements of an open and shorted balun.  

We initially estimate the ground loss from our beam model computed with FEKO (Section~\ref{section_beam_chromaticity}). Due to the simulation complexities and uncertainties in this small effect, we conduct an additional simulation with the CST package\footnote{\url{www.cst.com}}. The FEKO simulation employs Greens functions to model our finite metal ground plane over an infinite soil (infinite in $\pm\hat{x}$, $\pm\hat{y}$, and toward $-\hat{z}$). The soil is characterized in terms of its conductivity and relative permittivity, with nominal values of $0.02$ Sm$^{-1}$ and $3.5$, respectively, as estimated by \citet{sutinjo2015} for the dry conditions at the MRO. The CST simulation incorporates the ground plane but not the soil, representing a boundary scenario for the estimation of the beam fraction below the horizon. The two simulations produce an average loss of $0.5\%$ across the band, but with a different spectral profile. Therefore, we take as our nominal ground loss a spectrally flat profile with a value of $0.5\%$. Figure \ref{figure_antenna_losses_s11} (b) shows our fiducial antenna and ground losses.

\subsubsection{Beam Chromaticity}
\label{section_beam_chromaticity}

Beam chromaticity describes spectral variations of the antenna beam that introduce structure to the sky temperature spectrum \citep{vedantham2014, bernardi2015, mozdzen2016}. The calibrated sky temperature spectrum, $T_{\text{sky}}$, is obtained by removing the effect of beam chromaticity from the lossless antenna temperature, $T_{\text{ant}}$. We model this effect through a multiplicative chromaticity factor $C$ that relates the two temperatures as 

\begin{equation}
T_{\text{ant}}(\text{LST}, \nu) ~=~ C(\text{LST}, \nu) \cdot T_{\text{sky}}(\text{LST}, \nu).
\end{equation}

The chromaticity factor is computed as

\begin{equation}
C(\text{LST}, \nu) = \frac{\int \hat{B}(\nu, \Omega) \hat{T}_{\text{sky}}(\text{LST}, \nu, \Omega)d\Omega}{\int \hat{B}(\nu_n, \Omega) \hat{T}_{\text{sky}}(\text{LST}, \nu, \Omega)d\Omega},
\label{equation_chromaticity_factor}
\end{equation}

\begin{figure}
\centering
\includegraphics[width=0.49\textwidth]{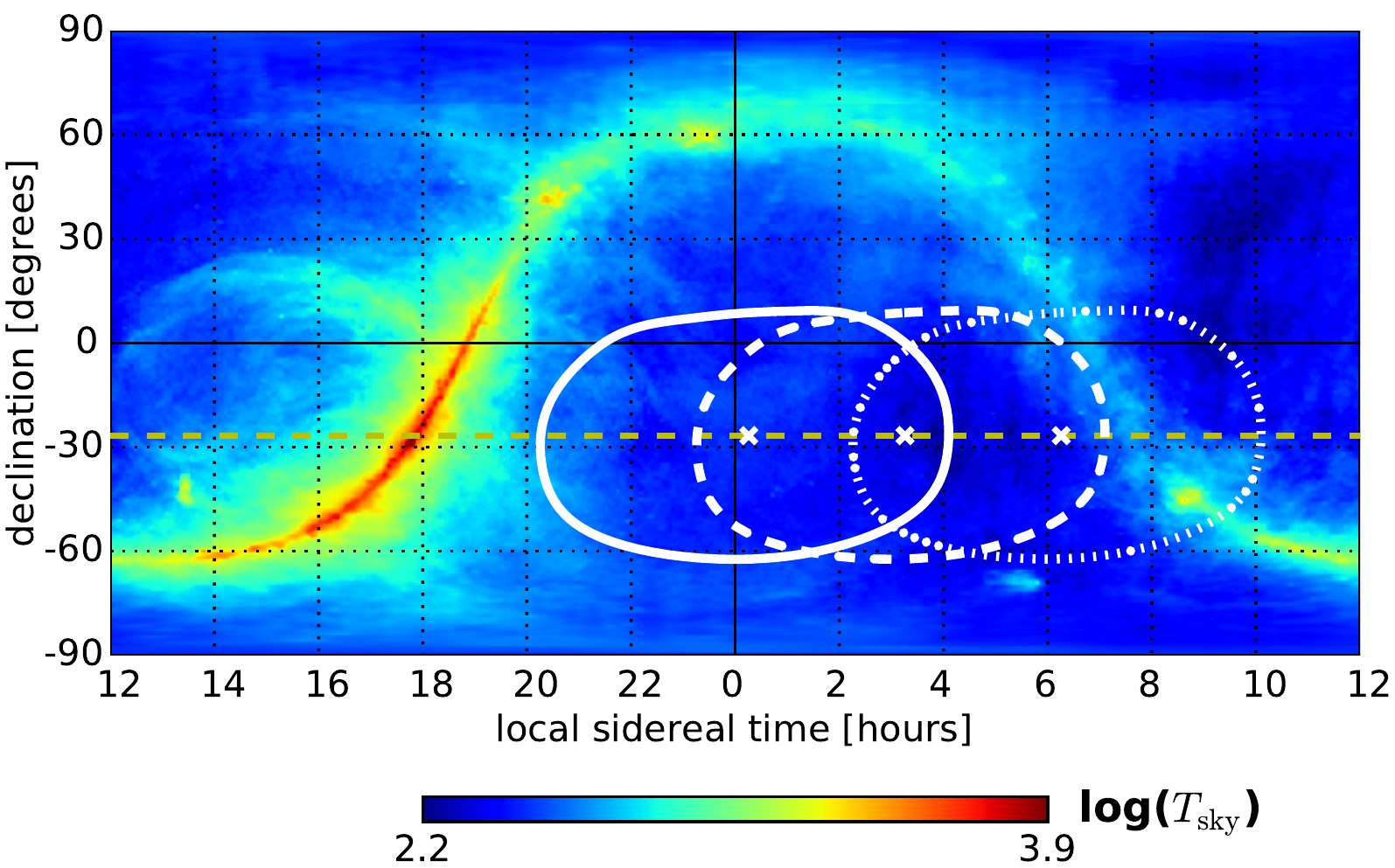}
\caption{Three reference projections of the $140$ MHz EDGES antenna beam FWHM onto the low foreground sky seen from the EDGES latitude of $-26.72^{\circ}$ (dashed yellow line). In this paper, we use observations over the continuous LST range $0.26-6.26$ hr. The three beam snapshots correspond to pointings at LST = $0.26$ hr (solid white), $3.26$ hr (dashed white), and $6.26$ hr (dotted white).}
\label{figure_scan_region}
\end{figure}

\noindent where $\Omega$ represents spatial coordinates above the horizon, $\hat{B}$ represents our beam model, and $\hat{T}_{\text{sky}}$ is a model for the diffuse sky visible from the MRO (different from our calibrated sky temperature spectrum, $T_{\text{sky}}$).

The numerator in Equation (\ref{equation_chromaticity_factor}) corresponds to the convolution of the frequency-dependent beam and sky, while the denominator represents a convolution where the beam is evaluated at a specific frequency $\nu_n$. Thus, by design, the denominator normalizes $C$ to one at $\nu=\nu_n$. At other frequencies, $C$ departs from one due to a beam pattern that changes with frequency. The normalization frequency is chosen as the middle of the band, i.e., $140$ MHz. The chromaticity factor only removes the corruption introduced by a frequency-dependent beam. It does not intend to force a match between the calibrated sky spectrum, $T_{\text{sky}}$, and the spectrum predicted from the sky model. 

We obtain our beam model from electromagnetic simulations with FEKO, which are also used to estimate some of the loss terms (Section~\ref{section_antenna_losses}). Our sky model consists of the $408$~MHz Haslam map \citep{haslam1982} scaled to the range $90-190$ MHz using a spatially dependent spectral index computed using the Haslam map and the Guzm\'an map at $45$ MHz \citep{guzman2011}. In \citet{mozdzen2017} we found that this simple model produces the closest match to our observed diffuse foreground spectra. The sky model does not include the global $21$ cm signal, which has an insignificant effect on the chromaticity correction.

Figure \ref{figure_scan_region} shows the FWHM of the beam model projected onto the sky model at $140$ MHz.

\begin{figure}
\centering
\includegraphics[width=0.48\textwidth]{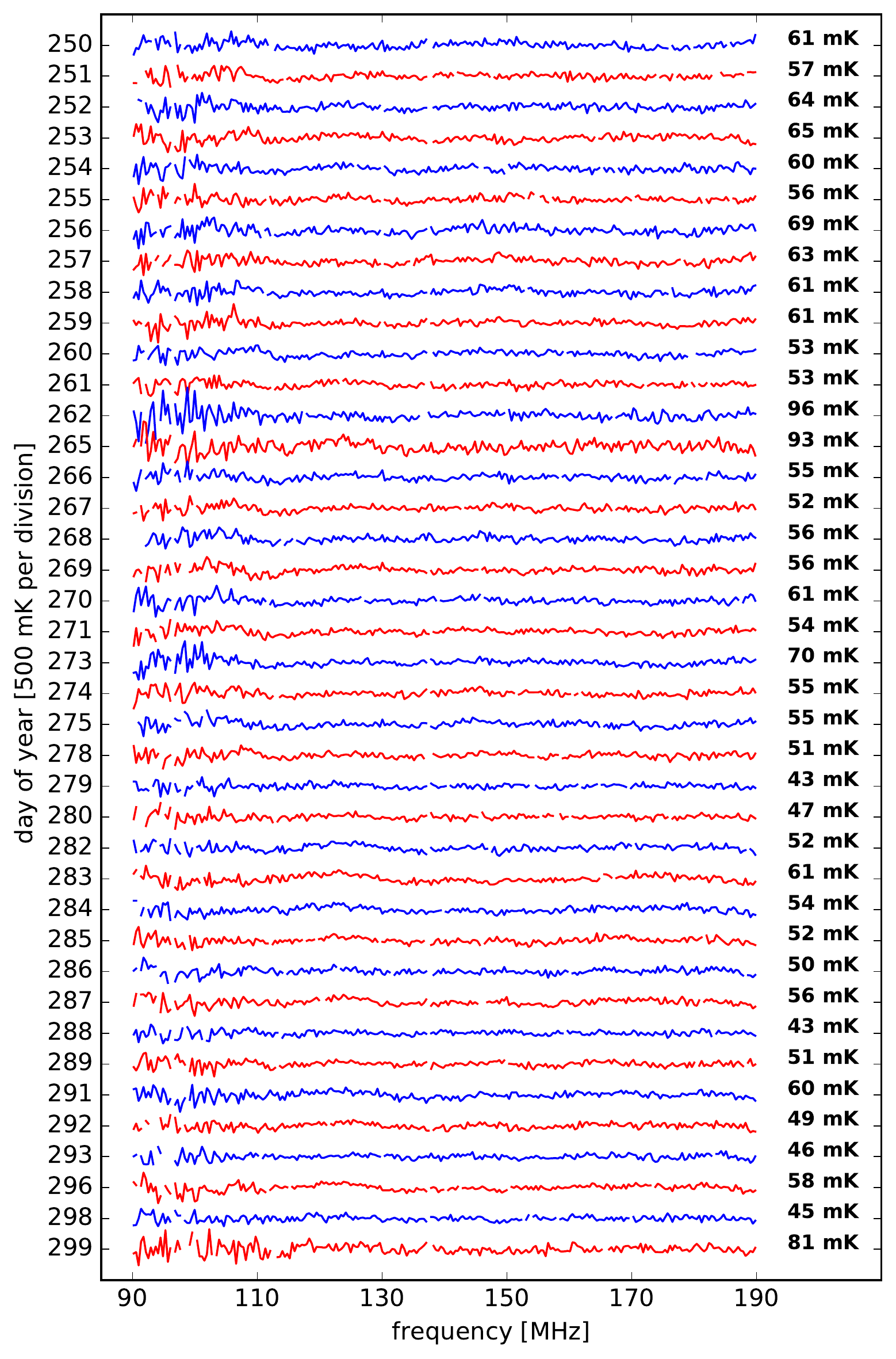}
\caption{Residuals of integrated spectra after five-term polynomial fits.  As described in Section~\ref{section_calibrated_spectrum}, in our analysis we use nighttime observations conducted between September 7 and October 26, 2015, over the LST range $0.26-6.26$ hr.  Each row represents the residuals of the integrated spectrum for a different night.  The rms of each residual spectrum is listed to the right and calculated with a frequency binning of $390.6$ kHz and weighted by the number of samples per bin.  The residual rms for each night varies between $43$ and $96$ mK. }
\label{figure_daily_residuals}
\end{figure}

\begin{figure}
\centering
\includegraphics[width=0.48\textwidth]{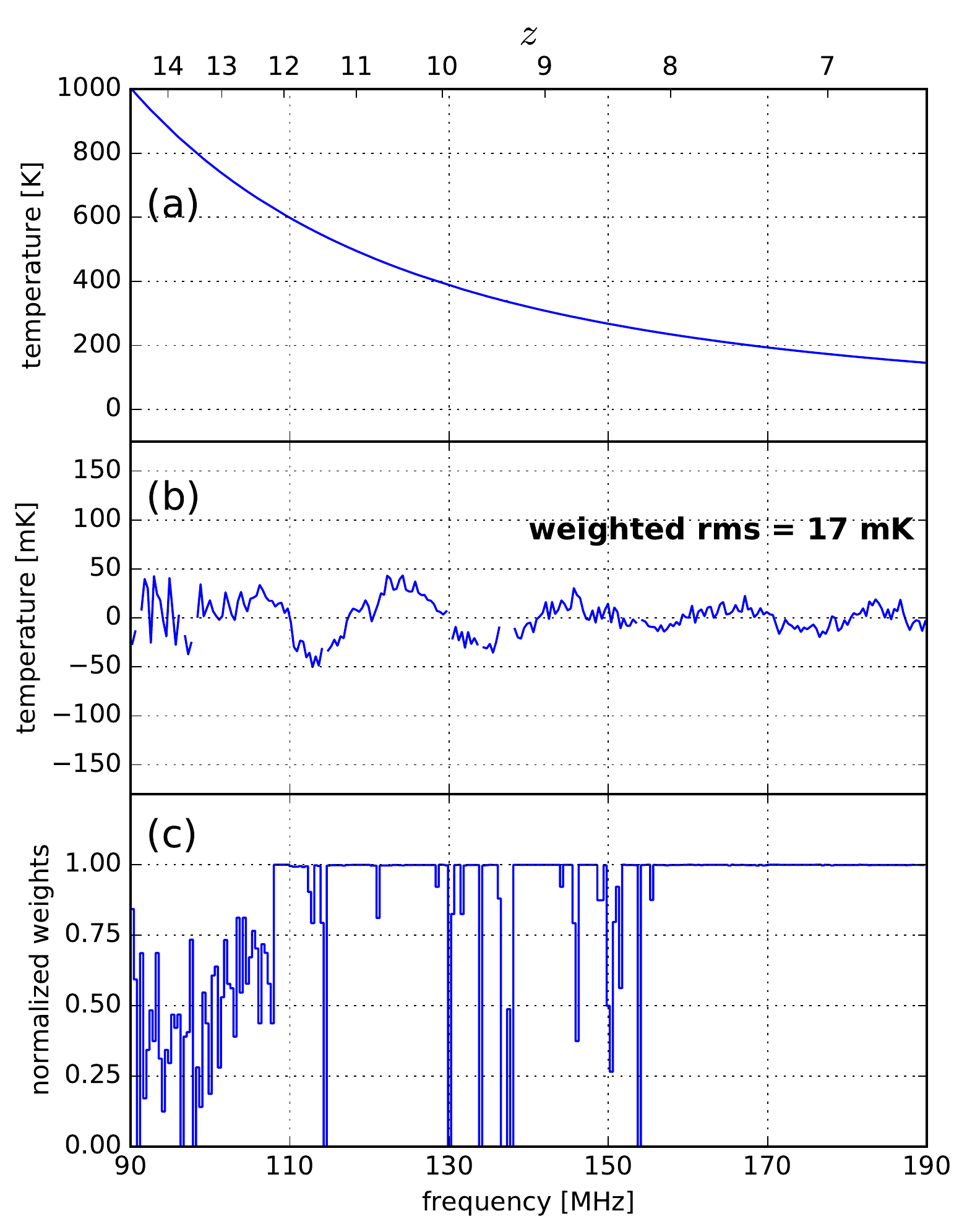}
\caption{(a) Total spectrum, integrated over LST $0.26-6.26$ hr. (b) Residuals of the integrated spectrum to a five-term polynomial fit. The weighted rms is $17$~mK. The rms of thermal noise alone above 108~MHz is $6$~mK.  Much of the observed ripple structure is non-thermal. (c) Normalized weights for each $390.6$~kHz spectral channel bin in the full integrated spectrum.  The weights are the fraction of raw samples in each bin compared to the maximum number of raw samples per bin, of approximately $1.2\times10^6$.  Variations in weight from channel to channel are due primarily to RFI excision, which is most pronounced in the FM radio band (below 108~MHz).}
\label{figure_final_residuals}
\end{figure}

\section{Data Analysis}
\label{section_data_analysis}

In this section we describe the data and our strategy for probing the $21$ cm models. 

\subsection{Calibrated Spectrum} 
\label{section_calibrated_spectrum}

The data used in this analysis correspond to $40$ nighttime observations, with the Sun at least $10^{\circ}$ below the horizon, covering the LST range $0.26-6.26$ hr. On the first day in our set, September 7, the coverage is only $0.26-4.5$ hr, which increases gradually and reaches $0.26-6.26$ hr on October 14. Since then, it remains constant until our last day, October 26.

We start the analysis by evaluating the structure and stability of the daily average spectra. For each daily data set, we (1) calibrate all $39$-second spectra, (2) average the spectra over LST $0.26-6.26$ hr, and (3)  bin the average spectrum into $390.6$ kHz channels ($64$ raw channels of $6.1$ kHz width). In parallel, we excise raw channels contaminated with RFI.  RFI is identified by averaging data on different time scales, between the shortest ($39$ s) and the daily ($>4$ hr) integrations, and performed both, before and after spectral binning. 

Then, to evaluate the quality and stability of the calibration we remove the contribution of diffuse foregrounds from each daily spectrum. Although to first order the foregrounds are well modeled by a power law \citep{mozdzen2017}, to reach the mK level targeted for the $21$ cm measurement the model requires additional terms \citep{kogut2012, voytek2014, bernardi2015, bernardi2016, sathy2017}. We find that five polynomial terms are necessary to reduce the fit residuals to our daily noise level over the $90-190$ MHz range. Our polynomial foreground model is: 

 \begin{equation}
 \hat{T}_{\text{fg}}(\nu) = \sum_{i=0}^4a_i\nu^{-2.5+i}.
 \label{equation_EDGES_polynomial}
 \end{equation}

Figure \ref{figure_daily_residuals} shows the single-day residuals to the five-term polynomial fit, as well as their weighted rms over frequency. In most cases the residuals are dominated by noise. Their rms fluctuates between $43$ mK and $96$ mK, with a median of $56$ mK. The day-to-day variations are attributed to (1) intrinsic noise variance, (2) different effective integration time due to different LST coverages, RFI excision, or bad weather cuts, and (3) any unaccounted for variations in instrument response not corrected in the calibration.

Next, we evaluate the structure and stability of the spectrum for longer integrations. We perform many trials of 20-day averages (about 50\% of total data), selecting days randomly for each trial.  We see that the residual structure to a five-term polynomial fit is consistent between the trials, with rms differences $<4$~mK.

Finally, we compute the total average spectrum. We average in time all the daily measurements at raw frequency resolution, and then bin the spectrum at $390.6$ kHz. In parallel we conduct the RFI excision, which is extended to time scales between several hours and the full set. The largest number of raw samples per frequency bin in the final spectrum is approximately $1.2\times10^6$ (with each sample representing $39$ s, $6.1$ kHz). 

Figure \ref{figure_final_residuals} (a) shows the total average spectrum, which has a brightness temperature $\approx 998$ K at $90$ MHz and $\approx 146$ K at $190$ MHz. Figure \ref{figure_final_residuals} (b) shows the residuals of the average spectrum to a five-term polynomial fit, and Figure \ref{figure_final_residuals} (c) shows the normalized number of samples per bin. The residuals have a weighted rms of $17$~mK over $90-190$ MHz.  Above 108~MHz, the rms of the thermal noise alone is $6$ mK, computed from the channel-to-channel differences.  Below 108~MHz, the thermal noise is higher due increased RFI excision in the FM radio band.  The non-thermal systematic structure can be described as ripples with a period of $\approx 20$ MHz and an amplitude that decreases with frequency. This pattern resembles the effect of small errors in the reflection coefficient of the receiver and antenna, as explored in \citet{monsalve2017}.  Our analysis described below accounts for systematic uncertainties and is designed to reduce sensitivity to any residual calibration errors in the spectrum by performing parameter estimation trials on the spectrum within windows of different widths and centers, as well as by using different numbers of foreground terms.  We will further explore the effects of possible calibration errors on our results in Section~\ref{section_calibration_uncertainties}.

\subsection{Model Rejection Approach}
\label{section_rejection_approach}

Following \citet{bowman2010}, we will report $21$~cm models that can be rejected at a given significance. We start by modeling the sky brightness temperature spectrum as:

\begin{equation}
T_{\text{sky}} = \hat{T}_{21} + \hat{T}_{\text{fg}} + \text{noise},
\end{equation}

\noindent where $T_{\text{sky}}$ is the observed spectrum data, $\hat{T}_{21}$ is one of the $21$ cm models, described in Section~\ref{section_results}, and $\hat{T}_{\text{fg}}$ is the foreground model of Equation (\ref{equation_EDGES_polynomial}).

The fit parameters in this model are the amplitude of the $21$ cm model, $a_{21}$, and the polynomial coefficients of the foreground model, $a_i$. We estimate the vector of linear parameters, $\lambda=[a_{21}, a_i]$, and their covariance matrix, $\Sigma$, using weighted least squares:

\begin{equation}
\hat{\lambda} = \left(A^TWA\right)^{-1}A^TWT_{\text{sky}},
\label{equation_estimates}
\end{equation}

\begin{equation}
\hat{\Sigma} = s^2\left(A^TWA\right)^{-1}.
\label{equation_covariance_matrix}
\end{equation}

Here, A is the design matrix, with columns that correspond to the normalized $21$ cm model and the polynomial terms of the foreground model, and W is a diagonal matrix of relative weights, with diagonal elements equal to the number of samples per frequency bin. The $s^2$ factor is the weighted sum of squared residuals normalized by the degrees of freedom,

\begin{equation} 
s^2  = \frac{r^TWr}{N_{\nu} - N_{\lambda} - 1},
\end{equation}

\noindent where $N_{\nu}$ is the number of frequency bins, $N_{\lambda}$ is the number of parameters, and $r$ is the difference between the data and the best-fit model.

To determine if a $21$~cm model can be rejected, we evaluate the consistency of the least squares amplitude estimate for the model, $\hat{a}_{21}$, with zero and with the expected model amplitude, $T_{\text{ref}}$. We reject the model if the amplitude estimate satisfies $|\hat{a}_{21}| < T_{\text{ref}}/2$.   We further limit the set of rejected models reported here to only those that can be detected when they are artificially injected into the data.  All rejections presented in this paper satisfy these conditions. The significance of the rejection is given by:

\begin{equation}
\text{rejection significance} = \frac{T_{\text{ref}} - \hat{a}_{21}}{\hat{\sigma}_{21}},
\label{equation_significance}
\end{equation}
where $\hat{\sigma}_{21}=\hat{\Sigma}_{11}^{1/2}$ is the least squares amplitude uncertainty estimate from the model fit.

We report rejections with significance equal to or higher than $1\sigma$. Models could fail to be rejected for two reasons. First, the fit amplitude, $\hat{a}_{21}$, could be far from zero (i.e., $|\hat{a}_{21}| \geq T_{\text{ref}}/2$) because the $21$~cm model fits structure in the integrated spectrum from either an actual $21$~cm signal or calibration errors. Second, the fit uncertainty, $\hat{\sigma}_{21}$, could be large due to high residuals from thermal noise or residual structure from using an imperfect signal model, or, due to high covariance between the $21$~cm and foreground models, which is likely, for example, with slow reionization scenarios. In particular, when we simultaneously fit the parameters $\lambda=[a_{21}, a_i]$, the estimate $\hat{a}_{21}$ does not systematically decrease as we increase the number of polynomial terms. Instead, it takes values consistent with larger uncertainty due to higher covariance.

To maximize the sensitivity to the $21$ cm~model through minimizing the residuals and covariances, we adjust the frequency range and the number of foreground parameters in the fit.   We compute multiple least squares estimates of $\hat{a}_{21}$ and $\hat{\sigma}_{21}$ for each trial 21~cm model by exploring multiple choices of the frequency range and foreground parameters.  We try (1) spectral windows of different widths between $20$ and $100$ MHz, (2) sweeping the window across the spectrum in steps equal to the frequency channel width ($390.6$ kHz), and (3) varying the the number of foreground terms between two and five. In the results below, we report the rejection significance obtained for the fit conditions that produce the lowest $21$ cm amplitude uncertainty.

\section{Results}
\label{section_results}

\begin{figure}
\centering
\includegraphics[width=0.48\textwidth]{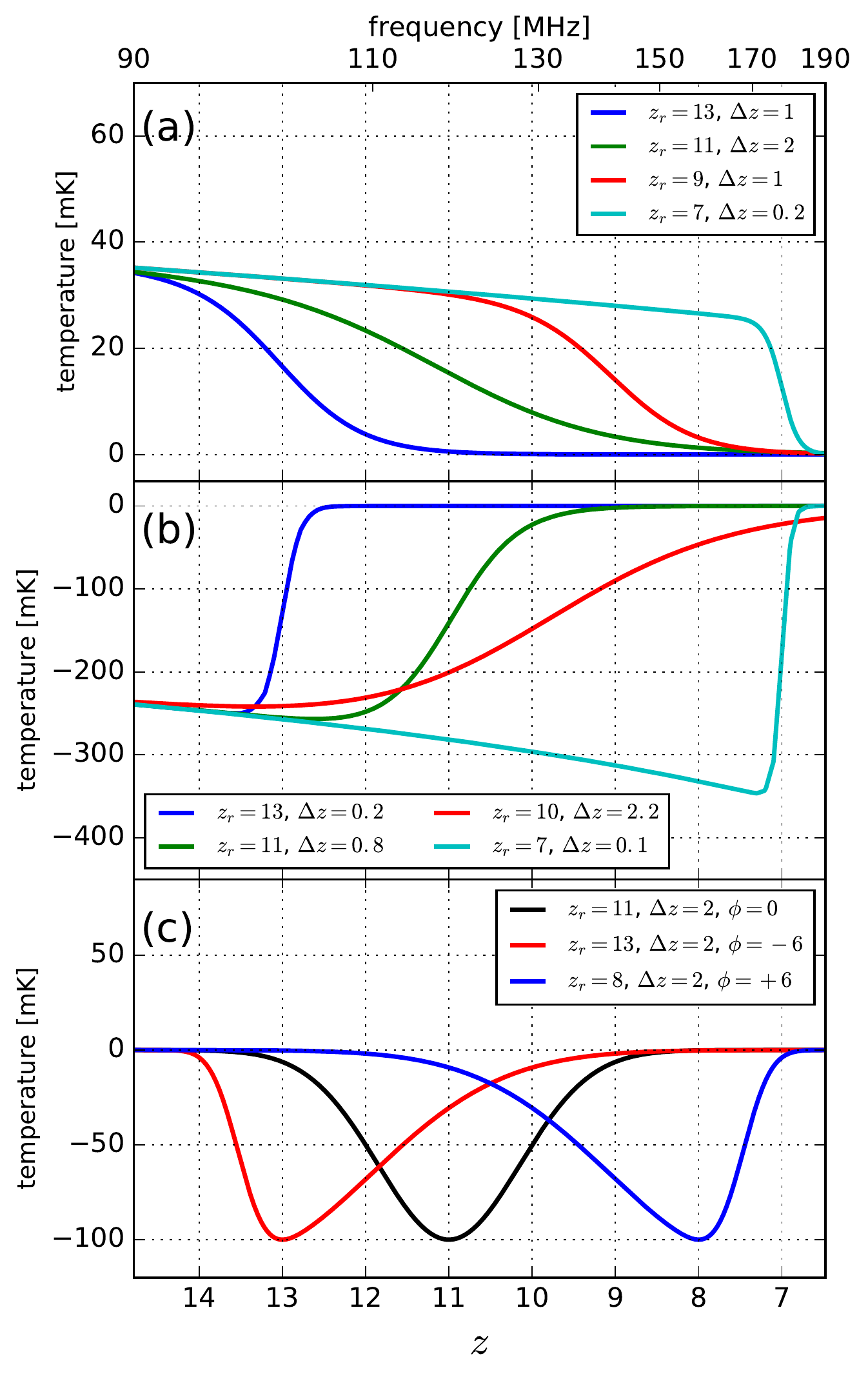}
\caption{Sample of phenomenological models representing the brightness temperature of the global $21$~cm signal. (a)~The top panel shows four models illustrating `hot' IGM reionization scenarios that assume effecient heating so that $T_s \gg T_{\text{cmb}}$ and use a tanh-based form for the evolution of neutral fraction, $x_{\text{HI}}$.  (b)~The middle panel shows models illustrating `cold' IGM reionization scenarios that assume no heating and also use a tanh-based form for $x_{\text{HI}}$. (c)~The bottom panel shows symmetrical and skewed Gaussian models for the absorption trough feature expected during the First Light era. In the Gaussian models, $\Delta z$ represents the FWHM of the Gaussian.}
\label{figure_phenomenological_models}
\end{figure}

Here we describe our findings for the three sets of $21$ cm models tested. They correspond to phenomenological models for the reionization transition assuming the two end-member cases for the heating states of the IGM, as well as a generic model for the absorption trough from the First Light era. Figure \ref{figure_phenomenological_models} shows a sample of these models. 

\begin{figure*}
\centering
\includegraphics[width=0.75\textwidth]{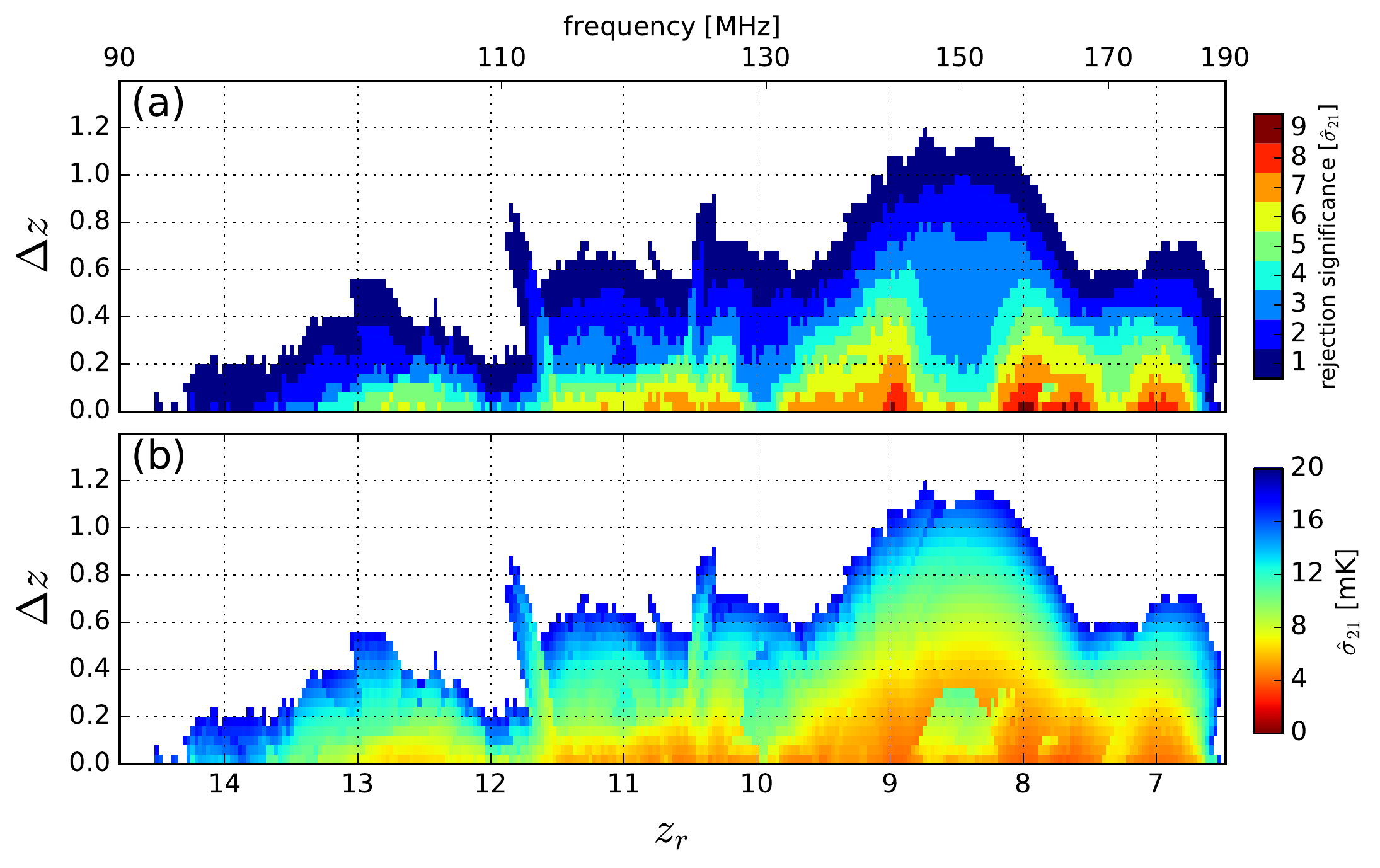}
\caption{Constraints on reionization from EDGES High-Band assuming a hot IGM and a tanh-based form for the evolution of the neutral fraction, $x_{\text{HI}}$. The reference amplitude of the 21~cm model is $T_{\text{ref}}=28$ mK. Parameters $z_r$ and $\Delta z$ indicate the redshift at 50\% reionization and the duration of reionization, respectively.  Colored regions represent models that are ruled out.  (a)~The top panel shows the model rejection significance with the color indicating discrete multiples of $\hat{\sigma}_{21}$.  (b)~The bottom panel shows the corresponding values of $\hat{\sigma}_{21}$. We reject models over the range $14.4\geq z_r\geq 6.6$, with a peak $2\sigma$ rejection of $\Delta z=1.0$ at $z_r\approx8.5$.}
\label{figure_hot_IGM_tanh}
\end{figure*}

\begin{figure*}
\centering
\includegraphics[width=0.99\textwidth]{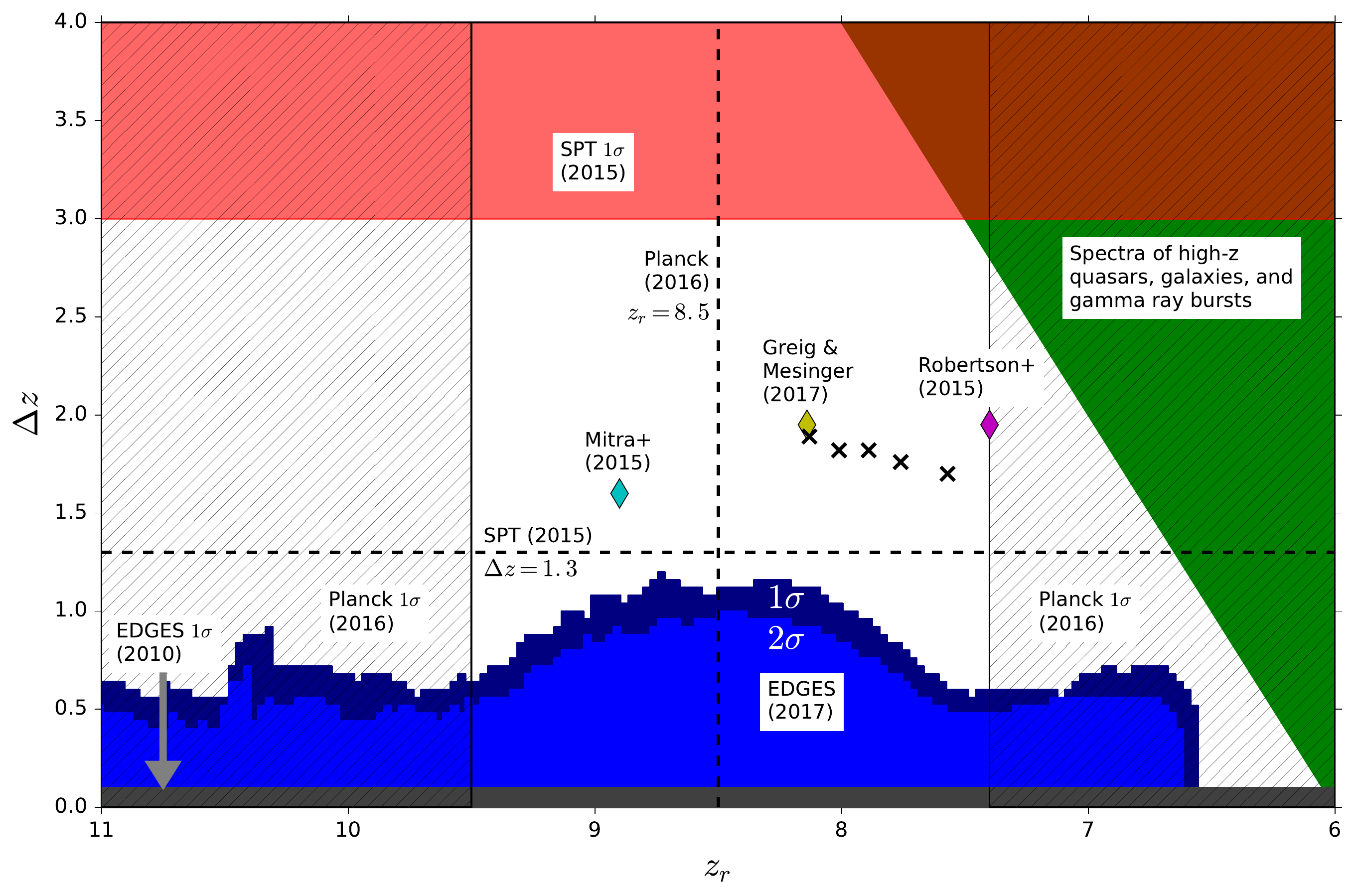}
\caption{Summary of reionization constraints from existing astrophysical observations.  Under the standard assumption of a hot IGM, EDGES rules out reionization models across the range $14.4\geq z_r\geq 6.6$ (blue region). The peak $2\sigma$ rejection corresponds to a duration $\Delta z=1.0$ at $z_r\approx8.5$. SPT reported an estimate (horizontal dashed line) and an upper limit (red region) for $\Delta z$ from measurements of the kSZ power in the CMB power spectrum \citep{george2015}. They defined $\Delta z$ as the range over which the ionized fraction (i.e., $1-x_{\text{HI}}$) increases from 0.2 to 0.99, and assumed the CMB optical depth reported by \emph{WMAP}. The \emph{Planck} estimate (vertical dashed line) and limits (hatched regions) for $z_r$ shown here assume a redshift-symmetric tanh form for $x_{\text{HI}}$ and no prior for the end of reionization redshift \citep{planck2016}. We also show a conservative generic upper limit (green region) of the form $\Delta z<2(z_r-6)$ to represent constraints from the spectra of high-$z$ quasars, galaxies, and gamma ray bursts \citep{bouwens2015}. Finally, we show specific reionization estimates from \citet{robertson2015} (magenta diamond), \citet{mitra2015} (cyan diamond), and \citet{greig2017} (yellow diamond for their `gold sample' and black crosses for other cases they considered). EDGES provides a unique constraint, ruling out fast reionization scenarios.  Modest additional improvements in EDGES performance will make it possible to directly probe the current best estimates for reionization.}
\label{figure_combined_EoR_constraints}
\end{figure*}

\subsection{Hot IGM Reionization}
\label{section_hot_igm_eor}

As discussed earlier, many models for the 21~cm signal during reionization find that the IGM has been heated by X-rays to substantially above the CMB temperature (e.g., \citealt{pritchard2012, mesinger2013}).  It is also found that the spin temperature is well coupled to the gas temperature by reionization.  Hence, in many cases, it is sufficient to approximate the detailed astrophysical models by assuming $T_{\text{s}}\gg T_{\text{cmb}}$ \citep{furlanetto2006a, furlanetto2006b}.  We follow this convention and first model the 21~cm brightness temperature for the reionization transition assuming X-ray heating of the IGM before the beginning of reionization.  With this assumption, Equation (\ref{equation_global_signal}) yields a simplified brightness temperature model:
\begin{equation}
\hat{T}_{21}(z) = a_{21} x_{\text{HI}}(z) \sqrt{\frac{1+z}{10}}.
\label{equation_tanh}
\end{equation}
We complete the model by representing the average neutral hydrogen fraction, $x_{\text{HI}}$, with the popular redshift-symmetric tanh expression (e.g., \citealt{bowman2010, pritchard2010, morandi2012, liu2013, mirocha2015, harker2016}),
\begin{equation}
x_{\text{HI}}(z) = \frac{1}{2} \; \left[\tanh\left(\frac{z-z_r}{\Delta z}\right) + 1\right].
\label{equation_neutral_fraction}
\end{equation}
The parameters in this $x_{\text{HI}}$ model are the reference redshift for $50\%$ reionization, $z_r$, and the reionization duration, $\Delta z=(dx_{\text{HI}}/ dz)^{-1}|_{x_{\text{HI}}=0.5}$. A similar tanh model for $x_{\text{HI}}$ has been used by CMB experiments \citep{lewis2008, hinshaw2013, planck2016}. 

As described in Section~\ref{section_rejection_approach}, the only $21$ cm fit parameter in our analysis is the brightness temperature amplitude, $a_{21}$. To determine if a model can be rejected, we compare the amplitude estimate, $\hat{a}_{21}$,  with the model reference temperature shown in Equation~(\ref{equation_global_signal}) of $T_{\text{ref}}=28$~mK \citep{madau1997, furlanetto2006b}. We explore the ($z_r$, $\Delta z$) domain to constrain reionization by sweeping the values of these parameters over the ranges $14.8 \geq z_r\geq 6.5$ and $0\leq\Delta z\leq2$.

Figure \ref{figure_hot_IGM_tanh} presents the rejection region for the hot IGM reionization models. In the top panel, the colors represent rejections at different discrete significance levels, between $1\sigma$ and $9\sigma$. For example, rejections in the range $[1\sigma, 2\sigma[$ are presented as $1\sigma$. The bottom panel shows the corresponding amplitude uncertainties $\hat{\sigma}_{21}$. The uncertainties are lower than $18$ mK; specifically, for rejections at $\geq 2\sigma$ they are $\leq 14$ mK, and for the peak $9\sigma$ rejections they are $\approx 4$ mK. In all the rejections, $\hat{a}_{21}$ is consistent with zero to within $\pm2\hat{\sigma}_{21}$ and the injection test yields amplitude estimates $\hat{a}_{21}^{inj}>2\hat{\sigma}_{21}$. As evident in the figure, we reject hot IGM reionization models over the range $14.4 \geq z_r \geq 6.6$. The strongest rejections are obtained in the central part of the spectrum due to the low noise and structure. In particular, our best rejection occurs at a  redshift $z_r\approx8.5$ ($150$ MHz). Here, we reject models with duration $\Delta z \leq 1.0$ at $\geq 2\sigma$ significance. For $z_r=8.5$ and $\Delta z=1$, the rejection is obtained from an amplitude estimate $\hat{a}_{21} = -9.7\times 10^{-4} \pm 13$~mK, using a $65$ MHz data window and three foreground polynomial terms. For four and five polynomial terms, the uncertainties grow to $49$ and $59$ mK, respectively, reflecting increased covariance due to overfitting. We also rule out, at $\geq 2\sigma$, models with duration $\Delta z \leq 0.8$  over the range $9 \geq z_r \geq 8$ ($142-158$ MHz) and $\Delta z \leq 0.4$ over $11.7 \geq z_r \geq 6.7$ ($112-184$ MHz). These reionization constraints represent a significant improvement with respect to our previous results of \citet{bowman2010}.

In Figure \ref{figure_combined_EoR_constraints} we present our new results in the context of existing constraints and estimates for reionization. A series of observations suggest that reionization has completed by $z\approx 6$ \citep{bouwens2015}, including the Gunn-Peterson trough \citep{becker2001, fan2006} and the fraction and distribution of `dark pixels' or gaps in the spectra of high redshift quasars \citep{gallerani2006, mesinger2010, mcgreer2015}, the Ly$\alpha$ damping wing absorption by neutral hydrogen in the spectra of quasars and gamma ray bursts \citep{chornock2013, schroeder2013, greig2017b}, the decrease in Ly$\alpha$ emission by galaxies at $z>6$ \citep{tilvi2014, choudhury2015}, and the clustering of Ly$\alpha$ emitters \citep{ouchi2010}. In the figure, we represent this combined constraint as an upper limit of the form $\Delta z < 2(z_r-6)$. The South Pole Telescope (SPT), from measurements of the kinetic Sunyaev-Zel'dovich (kSZ) effect imprinted on the $\ell\approx3000$ angular scales of the CMB power spectrum, estimated a reionization duration of $\Delta z=1.3$ and the upper limit $\Delta z < 3$ at $68\%$ confidence \citep{george2015}. SPT defined the duration as the difference between the redshifts for ionized hydrogen fractions $0.2$ and $0.99$, and assumed the CMB optical depth ($\tau_e$) reported by \emph{WMAP} \citep{hinshaw2013}. \emph{Planck} recently estimated a reionization redshift $z_r = 8.5^{+1.0}_{-1.1}$ assuming a redshift-symmetric transition \citep{planck2016}. This is the result shown in the figure for reference. Other estimates from \emph{Planck} range from $z_r=8.8$, derived after applying the prior $z>6$ for the end of the EoR, to $z_r=7.2$, combining their temperature and polarization measurements with high-$\ell$ data from SPT \citep{george2015} and ACT \citep{das2014}. We also incorporate in the figure reionization estimates by \citet{robertson2015}, \citet{mitra2015}, and  \citet{greig2017}, which themselves have been derived from different combinations of constraints. In this context, EDGES uniquely contributes to reducing the allowed phenomenological parameter space for a hot IGM reionization by ruling out tanh-based models with duration $\Delta z < 1.2$ at significance levels between $1\sigma$ and $9\sigma$.

\subsection{Cold IGM Reionization}
\label{section_cold_igm_eor}

Next, we probe models produced under the other extreme assumption of no IGM heating by X-rays before or during reionization. This condition results in a large global $21$~cm signal observed only in absorption. In particular, the amplitude of 21~cm absorption signal follows the adiabatic cooling of the IGM until, at the low-redshift end, the trough ends purely due to the extinction of neutral hydrogen during reionization. 

The cold IGM models are generated analytically by assuming (1) perfect Ly$\alpha$ coupling at early times such that $T_s=T_{\text{IGM}}$ in our observed redshift range, and (2) no X-ray or other heating, thus the IGM continues to cool adiabatically throughout reionization.  As with the hot IGM scenario, in these models the reionization histories follow Equation~(\ref{equation_neutral_fraction}) where the only parameters are the redshift and duration, for which we sample the ranges $14 \geq z \geq 6$ and $0.1\leq \Delta z \leq 4$, respectively. The resulting CMB optical depth for these models covers the range $0.128\geq \tau_e \geq 0.038$, which is significantly broader than the range currently prefered by \emph{Planck}, of $0.07 \geq \tau_e \geq 0.046$ at $68\%$ confidence \citep{planck2016}. The models have absorption peaks in the range $\approx200-370$ mK. As before, the only $21$ cm fit parameter is the model amplitude.

\begin{figure}[t!]
\centering
\includegraphics[width=0.49\textwidth]{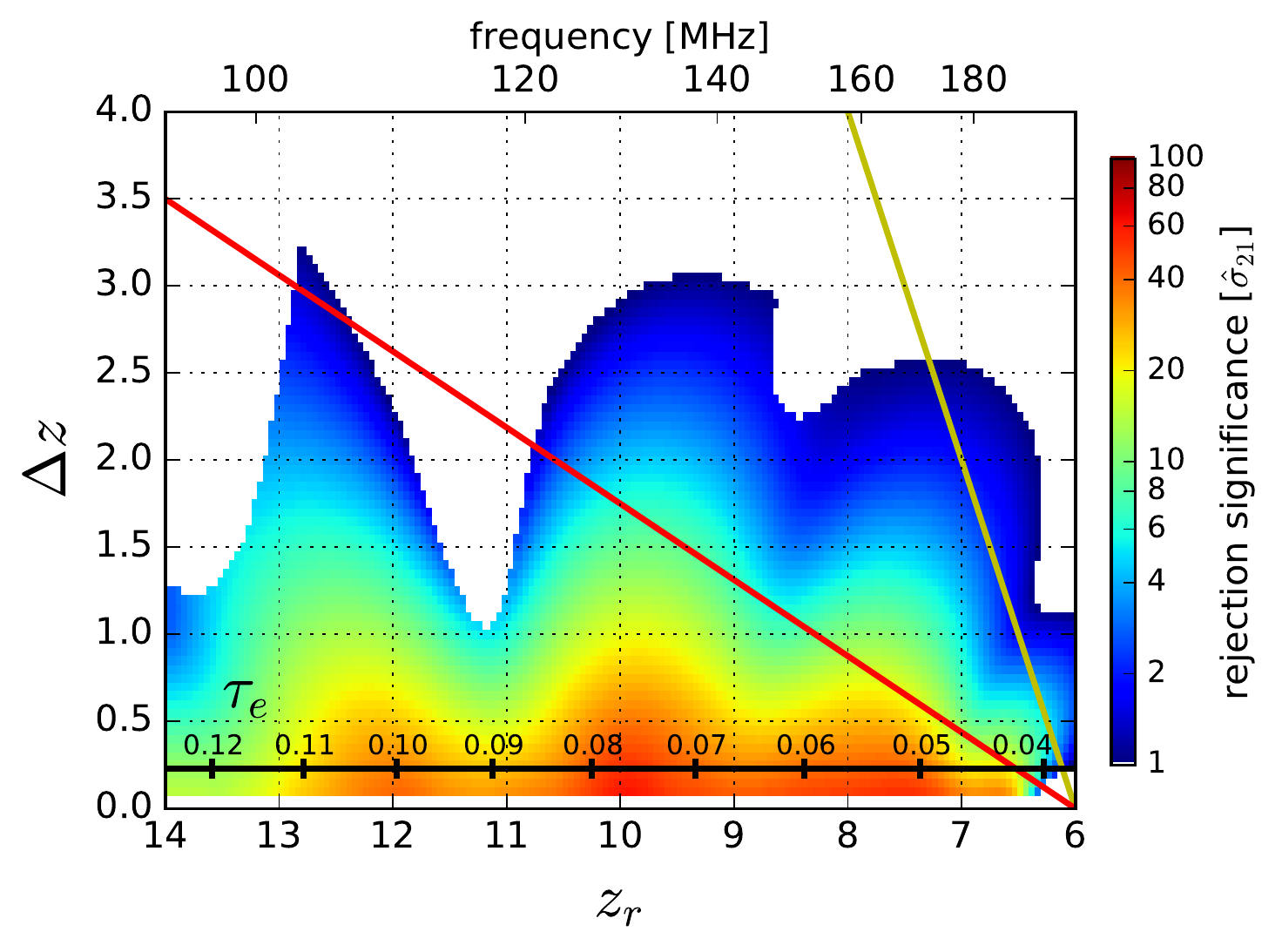}
\caption{Constraints on cold IGM reionization models from EDGES High-Band. These models assume perfect Ly$\alpha$ coupling at high redshifts and no IGM heating before or during reionization. The evolution of the neutral fraction, $x_{\text{HI}}$, follows the same tanh-based forms as in the hot IGM scenario. The rejections reach $\approx60\sigma$ significance for durations $\Delta z<0.5$. We show a horizontal reference scale for $\tau_e$ and two diagonal lines that represent upper limits on $\Delta z$ inferred from galaxy observations.  The yellow line shows the constraint of $\Delta z < 2(z_r-6)$ as described Section~\ref{section_hot_igm_eor}. The red line shows the more `aggressive' constraint of $x_{\text{HI}}<1\%$ at $z=6$. We rule out all of the probed cold IGM reionization models that satisfy $x_{\text{HI}}<1\%$ at $z=6$ and a CMB optical depth $0.086 \geq \tau_e \geq 0.038$, which approximately corresponds to the 2$\sigma$ limits from \emph{Planck}.}
\label{figure_cold_IGM_eor}
\end{figure}

Figure \ref{figure_cold_IGM_eor} presents the rejection results for the cold IGM reionization models. We reject a wide range of models, with durations of up to $\Delta z \approx3$ at $1\sigma$ significance, and significances of up to $\approx60\sigma$ for durations $\Delta z<0.5$. The average rejection limit across the probed range is $\Delta z \approx 2.2$. In the parameter estimation, the highest sensitivity to these models is achieved using a single $100$ MHz window (i.e., the full spectrum) and five polynomial terms.  The envelope of the rejection region shows the characteristic shape expected for a fit model consisting of a tanh reionization transition and a foreground polynomial with a fixed number of terms (see e.g., \citealt{pritchard2010, morandi2012}). The redshifts that most significantly depart from this pattern are $11.8\gtrsim z_r\gtrsim 10.7$ and $z_r\gtrsim 12.8$, which show constraints below the ripple envelope due to higher structure in the spectrum and higher noise, respectively, at the corresponding frequencies. In particular, the dip at $z_r\approx 11$ reduces the rejection limit to $\Delta z \approx1$ due to the relatively sharp feature in the spectrum at $110-125$ MHz.

Inside the figure, we show a horizontal scale with the values of $\tau_e$ which, for the range of $\Delta z$ rejected, are approximately proportional to $z_r$ and weakly dependent on $\Delta z$. For reference, we also show two diagonal lines. They represent upper limits on $\Delta z$ corresponding to (yellow) the constraint $\Delta z < 2(z_r-6)$ discussed in Section~\ref{section_hot_igm_eor}, and (red) a more `aggresive' constraint that limits the average hydrogen neutral fraction to $x_{\text{HI}}<1\%$ at $z=6$, consistent with some of the measurements of high-$z$ quasar spectra introduced in Section~\ref{section_hot_igm_eor} (e.g., \citealt{fan2006}).  Reionization scenarios that fall above and to the right of these lines are not allowed by these reference upper limits.  

We reject at $\geq 2\sigma$ all models that satisfy $0.086 \geq \tau_e \geq 0.038$ and $x_{\text{HI}}<1\%$ at $z=6$. Since this $\tau_e$ range approximately matches the $\pm2\sigma$ range reported by \emph{Planck}, we conclude that a scenario with perfect Ly$\alpha$ coupling at early times and no IGM heating before or during reionization is strongly inconsistent with current constraints on $\tau_e$ and $x_{\text{HI}}$. The strong rejection of these extreme models is, on the other hand, consistent with most theoretical expectations and with the lower limits on the $21$~cm spin temperature published by PAPER from interferometric measurements \citep{jacobs2015, ali2015, pober2015, greig2016}.

\subsection{Gaussian Absorption Features}

Finally, we probe a broad range of Gaussian models for the global 21~cm absorption trough expected during the First Light era.  As discussed above for the cold IGM reionization models, the 21~cm signal is predicted to become visible as an absorption feature against the CMB when Ly$\alpha$ photons from early stars couple the 21~cm spin temperature to the gas temperature prior to any heating.   However, in this more general scenario, the absorption signal ends due to a combination of the eventual X-ray heating of the IGM expected from stellar remnants and, in some cases, reionization.  This differs from the cold IGM reionization models that assumed the signal ending solely from reionization.  With the current lack of significant constraints, Gaussian shapes continue to be reasonable, generic models for the absorption trough in simulations and analysis of data \citep{bernardi2015, presley2015, bernardi2016}. Although some recent physical models depart from Gaussian shapes and exhibit sharper features (e.g., \citealt{fialkov2016a, cohen2017}), several models that predict the trough in our observed band ($\geq 90$ MHz) resemble a Gaussian to first order (e.g., \citealt{sitwell2014, kaurov2016, mirocha2017}).

Our test models correspond to Gaussians in redshift, and incorporate a skewness term to account for redshift asymmetry:

\begin{align}
\hat{T}_{21}(z) &= -a_{21} \; \exp\left[{-(4\ln2)\left(\frac{z-z_r-k_1}{\Delta z - k_2}\right)^2}\right] \times \nonumber \\
&\;\;\;\;\;\;\;\;\;\Bigg\{1 + \text{erf}\left[ \sqrt{4\ln2} \left( \frac{z-z_r-k_1}{\Delta z - k_2} \right) \phi\right]\Bigg\}.
\label{equation_gaussian}
\end{align}

The skewness term is based on the error function (erf) and the skewness parameter $\phi$. A symmetrical Gaussian is obtained by setting $\phi=0$. In this model, $a_{21}$ is the peak absorption amplitude, $z_r$ is the peak absorption redshift, and $\Delta z$ is the FWHM of the trough. The parameters $k_1$ and $k_2$ are used to compensate for the change in peak redshift and FWHM as $|\phi|$ increases for fixed $z_r$ and $\Delta z$. As with our previous models, the only $21$ cm fit parameter is $a_{21}$. For reference, a redshift symmetrical Gaussian with a FWHM in redshift of $\Delta z=4$ has a FWHM in frequency of $\approx 25.7$ MHz when centered at $z=14$ ($\approx 95$ MHz), and of $\approx 94.7$ MHz when centered at $z=7$ ($\approx 178$ MHz).

\begin{figure*}
\centering
\includegraphics[width=0.99\textwidth]{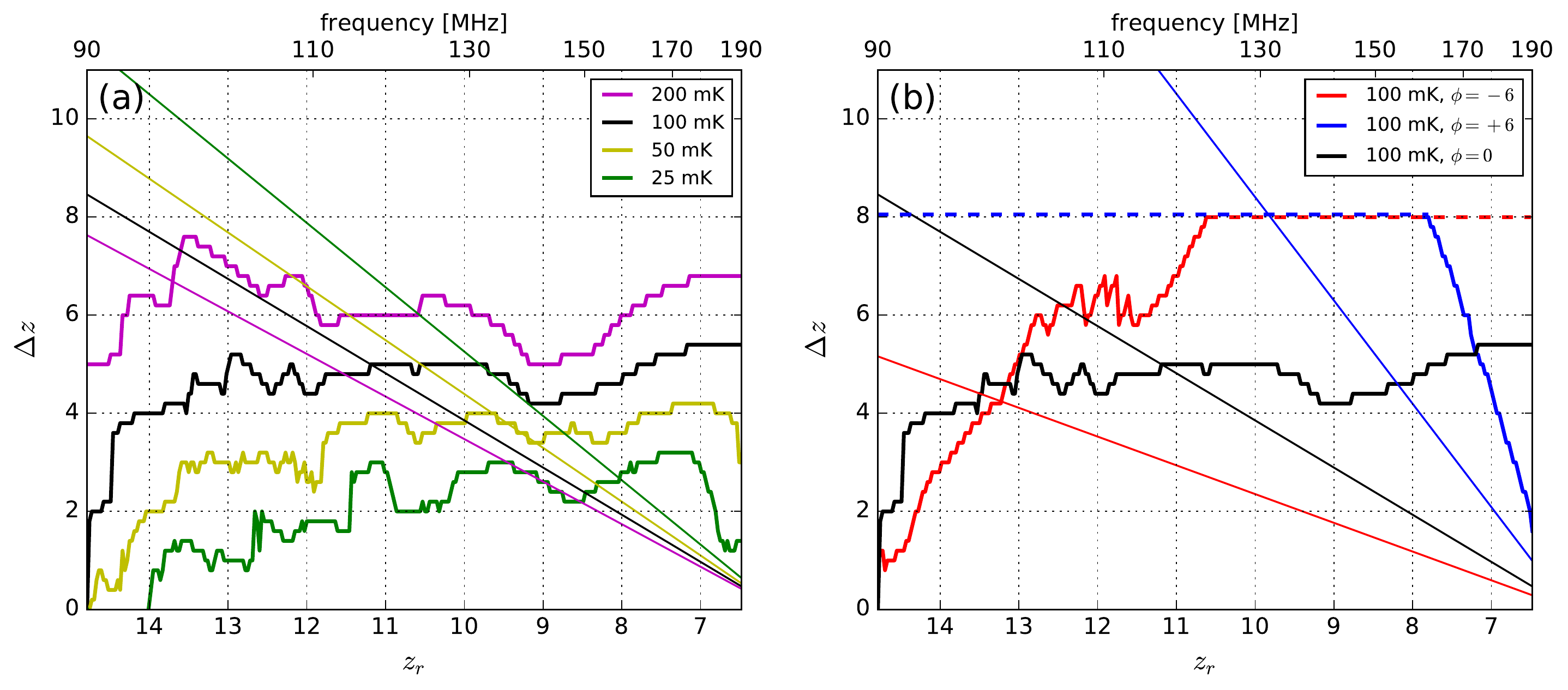}\\
\caption{Rejection limits at $2\sigma$ significance for Gaussian models of the 21~cm absorption trough expected in the First Light era. The thick lines are the limits derived from the EDGES data and the thin diagonal lines of the same color are reference upper limits for models consistent with a $21$~cm signal almost extinguished (brightness temperature between $-5$ mK and $0$ mK) at $z=6$. These reference upper limits are only shown to reflect that wide Gaussian-like models at low redshifts are inconsistent with scenarios where reionization is completed at $z\approx6$. (a) Limits for symmetrical Gaussians of amplitudes between $200$ and $25$ mK. We reject $200$~mK Gaussians with duration (FWHM) $\Delta z\leq5$ over $14.8\geq z_r\geq6.5$ and $25$~mK Gaussians with $\Delta z\leq2$ over $11.5 \geq z_r \geq 6.8$. (b) Limits for skewed $100$~mK Gaussians with $\phi=-6$ and $\phi=+6$, in addition to the symmetrical Gaussian reference. For both skewness values, the rejection limits grow above $\Delta z = 8$, but we conservatively report a maximum limit of $\Delta z=8$, depicted as dashed horizontal lines, because that is the total redshift width of our spectrum. We reject large fractions of the permitted regions below the diagonal limits, especially for amplitudes $\geq 100$~mK.}
\label{figure_results_gaussian}
\end{figure*}

Figure \ref{figure_results_gaussian} (a) shows the $2\sigma$ rejection limits for symmetrical Gaussians with reference amplitudes $T_{\text{ref}}=200$, $100$, $50$, and $25$ mK. The thick lines represent the results derived from the data, while the thin diagonal lines of the same color correspond to models that produce a brightness temperature of $-5$~mK at $z=6$. These diagonal lines illustrate our estimate of generic upper limits for models consistent with a $21$ cm signal almost extinguished by $z=6$.  From the EDGES observations, the following reference models can be rejected at $\geq2\sigma$: (1) for $200$ mK amplitude, models with duration $\Delta z \leq 5$ are rejected over the whole band, $14.8 \geq z_r \geq 6.5$; (2) for $100$ mK amplitude, models with duration $\Delta z \leq 4$ are rejected over $14.2 \geq z_r \geq 6.5$; (3) for $50$ mK amplitude, models with duration $\Delta z \leq 3$ are rejected over $11.8 \geq z_r \geq 6.5$; (4) for $25$ mK amplitude, models with duration $\Delta z \leq 2$ are rejected over $11.5 \geq z_r \geq 6.8$. Although not shown for simplicity, large fractions of models below these limits are rejected at $\geq9\sigma$ significance, especially for $100$ and $200$ mK amplitudes.

Independently, the reference diagonal upper limits rule out important regions of the ($z_r$, $\Delta z$) parameter space. With our data, we reject significant fractions below these limits. In particular, except for a small area above $z_r\approx13.7$, we rule out all $200$ mK Gaussian models in the allowed region.

Figure \ref{figure_results_gaussian} (b) shows $2\sigma$ rejection limits for skewed $100$~mK Gaussians with $\phi=-6$ and $\phi=+6$ (see Figure \ref{figure_phenomenological_models} (c)), and also includes, for reference, the symmetrical $100$~mK Gaussian limits. We show results for high skewness as clear examples of the effect of model redshift asymmetry on our rejection sensitivity. In particular, models with $\phi=+6$ resemble scenarios of rapid increase in the spin temperature due to strong heating, and/or a rapid decrease in the average hydrogen neutral fraction.

We reject asymmetric Gaussian models with $\phi=-6$ and $+6$ over larger widths than for the symmetrical Gaussian over most of the spectrum, since the asymmetric models have sharper features than the symmetric models for a given width and amplitude.  The asymmetric limits are lower than the symmetrical case only at the ends of the observed band.  For $\phi=-6$, rejections are low at high $z_r$ because only the smooth half of the model remains inside the spectrum. As the sharp half enters the spectrum and is swept to lower redshifts, our rejection sensitivity increases. The same occurs in the opposite direction for $\phi=+6$. The rejection limit increases above $\Delta z=8$ for both skewness values in opposite redshift directions. Because our spectrum has a total width of $\Delta z =14.8-6.5=8.3$, we conservatively report only a rejection limit capped at $\Delta z=8$ in these cases.  In Figure \ref{figure_results_gaussian} (b), we also show the diagonal upper limits for models that produce a brightness temperature of $-5$ mK at $z=6$. We see that only a small fraction of models with $\phi=-6$ are permitted by the limit, i.e., ones with durations $\Delta z\lesssim5$ at $z_r=14.8$, and our data rule out the majority of them. We rule out all the $\phi=+6$ models with $\Delta z\leq8$ in the region allowed under the limit.

\subsection{Calibration Uncertainties}
\label{section_calibration_uncertainties}

In the constraints presented above, we have assumed that calibration errors do not substantially affect the results. Here, we test that assumption. In order to estimate more broadly the effect of calibration errors on the results, we propagate uncertainties in the calibration parameters through our analysis pipeline following \citet{monsalve2017}. We perform the analysis described above in Section~\ref{section_data_analysis} and compute the $21$~cm rejection limits after calibrating the data using perturbed calibration parameter values. The perturbations are drawn randomly from Gaussian uncertainty distributions assigned to all relevant calibration parameters. The distributions are based on estimates of possible error levels and centered at the nominal calibration values. Table \ref{table_uncertainties} lists the calibration parameters and their $1\sigma$ widths. Running each calibration realization through our pipeline is computationally expensive. Therefore, we only conduct this process for the two $21$~cm models with the smallest amplitudes: the $28$~mK hot IGM reionization model and the $25$~mK symmetrical Gaussian absorption trough model. In both cases, we run $100$~realizations with all source uncertainties applied simultaneously. From these realizations, we derive first-order uncertainty estimates for our rejection limits.

\capstartfalse
\begin{deluxetable}{ll}
\tabletypesize{\scriptsize}
\tablewidth{0pt}
\tablecaption{Uncertainties Assigned to Calibration Parameters \label{table_uncertainties}}
\tablehead{\colhead{Parameter} & \colhead{$1\sigma$ Uncertainty}}
\startdata
\textbf{Receiver}                        & \\
Temperature correction                   & $0.1^{\circ}$C \\
Absolute calibration                     & from \citet{monsalve2017}\\
\\
\textbf{Antenna Reflection Coefficient}  & \\
Magnitude                                & $10^{-4}$ in voltage ratio (frequency rms)$^*$\\
Phase                                    & $0.1^{\circ}$ (frequency rms)$^*$\\
\\
\textbf{Antenna and Ground Losses}       & \\
Balun length                             & $1$ mm\\
Connector length                         & $0.1$ mm\\ 
Balun and connector radii                & $3\%$\\
Balun and connector conductivity         & $1\%$ \\
Connector teflon permittivity            & $1\%$ \\
Panel loss                               & $10\%$\\
Ground loss                              & $10\%$ of nominal + \\
                                         &$30\%$ from FEKO and CST$^{\dagger}$\\
\textbf{Chromaticity Factor}             & \\
Sky model                                & $50\%$ of difference between nominal\\ 
                                         & model and sky model from\\
                                         & \citet{zheng2017}\\
Antenna panel height                     & $2$ mm \\
Antenna panel length                     & $2$ mm \\
Antenna panel width                      & $2$ mm \\
Antenna panel separation                 & $1$ mm \\
Ground plane length                      & $5$ cm \\
Ground plane width                       & $5$ cm \\
Antenna orientation angle                & $0.5^{\circ}$ \\
Soil conductivity                        & $50\%$  \\
Soil relative permittivity               & $50\%$
\enddata
\tablecomments{Unless otherwise noted, percentages are given as relative to the nominal value.\\$^*$Perturbations modeled as polynomials in frequency, using a number of terms chosen randomly between $1$ and $16$ (same as for the nominal model), with frequency rms as listed in the Table. \\$^{\dagger}$Frequency-independent perturbations to the nominal value plus frequency-dependent perturbations proportional to the FEKO and CST results.}
\end{deluxetable}

We quantify the impact of the calibration uncertainties through the scatter of the $2\sigma$ limit on $\Delta z$. For the hot IGM models, the average standard deviation of this limit across frequency is $0.1$, with a peak of $0.13$ at the redshift of peak rejection, $z_r\approx8.5$. For the $25$~mK Gaussian absorption trough model, the average standard deviation is $0.27$. The $\sim 10\%$ variation in rejection limits observed in both cases reflects our low sensitivity to a combination of realistic calibration errors. We provide these estimates here as a complement to our nominal results. For future work, we will continue exploring techniques to understand the residual spectral structure in the integrated observation. 

Improvements in the instrument and calibration would result in a measurement that represents the sky spectrum with higher fidelity and with lower structure from systematics. This, in turn, could enable to reject more $21$ cm models, as well as establish stronger rejections for many of the currently rejected models due to the lower number of terms potentially required in the foreground model for a given frequency range.

\section{Conclusion}
\label{section_conclusion}

In this paper we evaluate the consistency between three sets of phenomenological models for the global $21$ cm signal and the average sky brightness temperature spectrum measured with EDGES High-Band between September $7$ and October $26$, $2015$. In summary, we derive the following constraints for each set of models:

\begin{enumerate}
\item We rule out various tanh-based models for reionization that assume the standard hot IGM  scenario ($T_s \gg T_{\text{cmb}}$) over the range $14.4 \geq z_r \geq 6.6$.   Our peak rejection at $2\sigma$ significance is $\Delta z = 1.0$ at $z_r\approx8.5$. We also reject models with duration $\Delta z \leq 0.8$ over $9 \geq z_r \geq 8$, and $\Delta z \leq 0.4$ over $11.7 \geq z_r \geq 6.7$. These new EDGES results represent a significant improvement with respect to \citet{bowman2010}, and directly complement the upper limits on the reionization duration from secondary CMB anisotropy measurements by SPT.  In addition, our peak rejection redshift occurs in the range of current \emph{Planck} estimates for the middle point of reionization.

\item For tanh-based models of reionization that assume an extreme cold IGM scenario with perfect Ly$\alpha$ coupling at early times ($T_s$ equal to the kinetic gas temperature) and no heating of the IGM before or during reionization, we reject at high significance (up to $\approx60\sigma$) all models that produce a CMB optical depth in the range $0.086 \geq \tau_e \geq 0.038$ and an average hydrogen neutral fraction $x_{\text{HI}}<1\%$ at $z=6$. The rejection of these models is consistent with the expectation of IGM heating during reionization.

\item We rule out a variety of generic Gaussian models for a $21$ cm absorption trough in our observed band. As a reference, over the observed spectrum we reject, at $\geq 2\sigma$ significance, redshift symmetrical Gaussians of amplitude $200$, $100$, $50$, and $25$ mK with durations $\Delta z \leq 5$, $4$, $3$, and $2$, respectively. For models with skewness, the rejection limit increases over most of the spectrum. In particular, when using $\phi=\pm6$ in our $100$ mK model, the rejection limit increases to $\Delta z=8$, which is approximately the full width of our measured spectrum.  
\end{enumerate}

We propagate our calibration uncertainties to estimate their simultaneous effect on our rejection results and find $\sim$~10\% scatter in the $2\sigma$ rejection limits for the most sensitive models. 

Future work is planned to build on this analysis, moving beyond phenomological models to directly probe astrophysical models for the global 21~cm signal using the EDGES High-Band data. Observations with the EDGES Low-Band system are presently being analyzed and are expected to extend the constraints reported here.  Ongoing instrument development is expected to continue to improve the performance of EDGES toward a goal of directly testing the current best estimates for the timing and evolution of reionization. 

\acknowledgements
We are grateful to the reviewer for useful suggestions to improve this paper. We also thank Jordan Mirocha for producing the cold IGM reionization models and for valuable feedback. We finally thank Rennan Barkana, Jack Burns, Steve Furlanetto, Adrian Liu, David Rapetti, Peter Sims, and Keith Tauscher for insightful discussions. This work was supported by the NSF through research awards for the Experiment to Detect the Global EoR Signature (AST-0905990, AST-1207761, and AST-1609450). Raul Monsalve acknowledges support from the NASA Ames Research Center (NNX16AF59G). EDGES is located at the Murchison Radio-astronomy Observatory. We acknowledge the Wajarri Yamatji people as the traditional owners of the Observatory site. We thank CSIRO for providing site infrastructure and support.

%
%
%

\clearpage

\end{document}